\documentclass[conference]{IEEEtran}
\IEEEoverridecommandlockouts
\usepackage{cite}
\usepackage{amsmath,amssymb,amsfonts}
\usepackage{algorithmic}
\usepackage{graphicx}
\usepackage{textcomp}
\def\BibTeX{{\rm B\kern-.05em{\sc i\kern-.025em b}\kern-.08em
    T\kern-.1667em\lower.7ex\hbox{E}\kern-.125emX}}
  \makeatletter
    \newcommand{\linebreakand}{%
      \end{@IEEEauthorhalign}
      \hfill\mbox{}\par
      \mbox{}\hfill\begin{@IEEEauthorhalign}
    }
    \makeatother

\usepackage{xcolor}

\begin{document}

\title{Mutual Reinforcement between Neural Networks and Quantum Physics
}

\author{\IEEEauthorblockN{1\textsuperscript{st} Yue Ban}
\IEEEauthorblockA{\textit{Department of Physical Chemistry } \\
\textit{University of the Basque Country UPV/EHU}\\
Bilbao, Spain \\
ybanxc@gmail.com}
\and
\IEEEauthorblockN{2\textsuperscript{nd} Javier Echanobe}
\IEEEauthorblockA{\textit{Department of Electricity and Electronics} \\
\textit{University of the Basque Country UPV/EHU}\\
Bilbao, Spain \\
franciscojavier.echanove@ehu.eus}
 \linebreakand 
\IEEEauthorblockN{3\textsuperscript{rd} Erik Torrontegui}
\IEEEauthorblockA{\textit{Departamento de F\'{i}sica} \\
\textit{Universidad Carlos III de Madrid}\\
\textit{Instituto de F\'{\i}sica Fundamental IFF-CSIC}\\ 
Madrid, Spain\\
eriktorrontegui@gmail.com}
\and
\IEEEauthorblockN{4\textsuperscript{th} Jorge Casanova}
\IEEEauthorblockA{\textit{Department of Physical Chemistry } \\
\textit{University of the Basque Country UPV/EHU}\\
\textit{IKERBASQUE,  Basque  Foundation  for  Science}\\
Bilbao, Spain \\
jcasanovamar@gmail.com}
}

\maketitle

\begin{abstract}
Quantum machine learning emerges from the symbiosis of quantum mechanics and machine learning. In particular, the latter gets displayed in quantum sciences as: (i) the use of classical machine learning as a tool applied to quantum physics problems, (ii) or the use of quantum resources such as superposition, entanglement, or quantum optimization protocols to enhance the performance of classification and regression tasks compare to their classical counterparts. This paper reviews examples in these two scenarios. On the one hand, a classical neural network is applied to design a new quantum sensing protocol. On the other hand, the design of a quantum neural network based on the dynamics of a quantum perceptron with the application of shortcuts to adiabaticity gives rise to a short operation time and robust performance. These examples demonstrate the mutual reinforcement of both neural networks and quantum physics. 
\end{abstract}


\section{Introduction}
Artificial neural networks \cite{McCulloch43} are one of the most fruitful computational models of machine learning (ML) \cite{ML} thanks to the blooming of deep learning \cite{deep-learning2,deep-learning3} in the recent years. Inspired by biological brains, artificial neural networks (NNs) are organized in layers feeding signals into other neurons allowing parallel-processed \cite{Hopfield84} and universal \cite{Cybenko89} computing. Introduced first in 1943 as a computational model based on discrete threshold logic algorithms \cite{McCulloch43} and linked to biological learning theory \cite{Hebb49}, the neuron activation mechanism --i.e. perceptron, was characterized  \cite{Rosenblatt58} and generalized  to graded response activation potentials \cite{Hopfield84}. When nested with other perceptrons, the resulting multilayer network becomes universal with the capacity to approximate any continuous function \cite{Cybenko89, Hornik89}, based on ``universal approximation theorem" claiming that any continuous function can be written as a linear combination of sigmoid functions.

The topology of NNs has increased in parallel with the computer hardware improvements \cite{Steinkrau05}. 
As a result, the enlargement of the depth has promoted the ability of a NN to process the exponentially increasing amount of data \cite{Walter05} and complexity of algorithms \cite{Kolmogorov63} in the information explosion era \cite{Castells96}. Thus, the versatility of NNs covers diverse fields such as economy, industry, transportation, or science among others.
More recently, the so-called deep learning paradigm 
and particularly the Convolutional Neural Networks (CNNs) \cite{CNN}, have shown incredible capabilities in applications such as speech \cite{Hinton12} or object recognition \cite{Hinton06}, spam filters \cite{Dada19}, vehicle control \cite{Devi20}.  
Neural networks can be trained to perform tasks without the programmer necessarily detailing how to do it. Due to this property, CNNs are also employed to develop the expected artificial intelligence.

Over the last two decades, quantum physics has experienced its second revolution giving rise to new quantum technologies. Thanks to quantum control, matter can be manipulated at the single particle level by exploiting quantum resources such as entanglement, superposition or squeezing of states in various platform registers with high fidelity \cite{quantum-platform1,quantum-platform2,quantum-platform3,quantum-platform4,quantum-platform5}.  All the progress indicates that quantum physics will offer outbreak for the future coming in a wide variety of forms, from quantum cryptography \cite{quantum-cryptography}, quantum sensing \cite{quantum-sensing}, to quantum computing \cite{quantum-computing} among others. Some examples of technologies emerged in this second quantum revolution are atomic sensors providing unprecedented high resolution and efficiency in the detection of external fields \cite{atomic-sensor1,atomic-sensor2,atomic-sensor3,atomic-sensor4,atomic-sensor5}, quantum  channels \cite{quantum-channel} that make use of entanglement for unbreakable communications \cite{quantum-communication}, or quantum computers that use superposition to execute parallel-processing computations performing specific tasks with higher efficiency than their classical counterparts \cite{google-supremacy}.

The meet of ML and quantum physics gives birth to a novel field, quantum machine learning (QML) \cite{QML1,QML2}, bringing a lot of progress on both fields. On the one hand, the universality of NNs allows to enhance the accuracy and efficiency of quantum protocols. ML is diversely useful in measurements protocols \cite{quantum-measurement1,quantum-measurement2,quantum-measurement3}, states preparation \cite{quantum-state1,quantum-state2}, entanglement and states classification \cite{classify-state2,classify-state3}, quantum communication  \cite{ML-quantum-communication}, learning Hamiltonians \cite{H-learning} and handling with open quantum system dynamics \cite{open-system-dynamics1,open-system-dynamics2,open-system-dynamics3}. Recently, ML also starts to attract its attention in quantum sensing and metrology, in particular, adaptive protocols for phase  \cite{ML-estimation1,ML-estimation3} and parameter estimation \cite{ML-estimation4,ML-estimation5}, or calibrating quantum sensors \cite{calibration-sensor}. On the other hand, the use of quantum resources allows improvements of the NN accuracy in classification and regression tasks compared to their classical counterparts
. Since the seminal publication by Kak \cite{Kak}, different efforts have been made to reproduce the nonlinear response for the perceptron at the quantum level and its nesting to the design of artificial quantum neural networks (QNNs) \cite{Mitarai18, Farhi18, Cao17, Perez20, reversible-circuit,Continuous-variable-QNN, quantum-perceptron, quantum-perceptron-STA, multi-NN}.  In particular, in Reference \cite{quantum-perceptron} it is already demonstrated that a QNN based on the quantum perceptron gates evolving adiabatically with the sigmoid-like activation has the ability to approximate continuous functions.

In this article, we review two examples of QML, demonstrating the uses of NNs in quantum physics. 
This article is structured as follows. In Sec. \ref{section:NN-sensing}, we review the use of a classical NN applied to quantum sensing \cite{NN-magnetometry}. Such a NN is used for parameter predictions of an external field, showing its performance in a wide range of working regimes, even beyond the rotating wave approximation (RWA) \cite{harmonic-response1,harmonic-response2} with a training/validation/test dataset fully obtained from experimental measurements. The establishment, training and operation of the NN require a minimal knowledge of the physical system in contrast with Bayesian inference methods \cite{atomic-sensor5}. In Sec. \ref{section:Q-perceptron}, we review the quantum perceptron where the sigmoid-like activation is mimic by the excitation probability of a qubit. Implemented as an efficient and reversible unitary operation, the resulting QNN is encoded by an Ising model \cite{quantum-perceptron}. With the aid of shortcuts to adiabaticity (STA) \cite{quantum-perceptron-STA} the perceptron can provide faster and more robust nonlinear responses. The results of these work indicate the mutual reinforcement of machine learning and quantum physics.

\section{Application of NNs in quantum sensing} \label{section:NN-sensing}
Target parameters have been usually encoded in harmonic responses of quantum sensors \cite{harmonic-response1,harmonic-response2}. However, this approach limits the working regime of the sensor, as the deviation from harmonic responses leads to a failure on the parameter estimation. In addition, being in this limited working regime  requires enough pre-knowledge of the physical model. In this scenario, a well-trained feed-forward NN (such as the one  proposed in \cite{NN-magnetometry}) can predict the parameters of the external field just from the experimental acquisitions without the necessity of learning the underlying physical model. In particular,  in the work conducted in \cite{NN-magnetometry} we focused on an $^{171}$Yb$^+$ atomic quantum sensor \cite{atomic-sensor1,atomic-sensor2,atomic-sensor3,atomic-sensor4,atomic-sensor5} and demonstrated that its working regime was extended to complex scenarios. 

The atomic sensor works as follows: In the $^2S_{\rm{\frac{1}{2}}}$ manifold of a $^{171}$Yb$^+$ ion, there are four hyperfine levels $|0\rangle$, $|\acute{0}\rangle$, $|1\rangle$ and $|-1\rangle$, with a Zeeman splitting in the late three states induced by a static magnetic field $B_z$. Two microwave drivings with the same amplitude $\Omega$ and resonant with $|0\rangle \rightarrow |1\rangle$ and $|0\rangle \rightarrow |-1\rangle$ are applied to cancel potential noisy fluctuations in $B_z$. Via the detection of the state transition between the $|\acute{0}\rangle \rightarrow |1\rangle$ (or $|\acute{0}\rangle \rightarrow |-1\rangle$), a target electromagnetic field with frequency $\omega_{\rm{tg}}$ and amplitude $\Omega_{\rm{tg}}$, in the form of $\Omega_{\rm{tg}} \cos(\omega_{\rm{tg}})$ can be probed based on the responses that the sensor gives. In the dressed state basis $\{|u\rangle, |d\rangle, |B\rangle, |D\rangle\}$, where $|u\rangle = (|B\rangle + |0\rangle)/ \sqrt{2}$, $|d\rangle = (|B\rangle-|0\rangle)/ \sqrt{2}$,  $|D\rangle = (|-1\rangle - |1\rangle)/\sqrt{2}$, $|\acute{0}\rangle = |\acute{0}\rangle$, with $|B\rangle = (|-1\rangle+|1\rangle)/\sqrt{2}$, the state probability $P_D(t)$ is considered as the response given by the sensor with $P_D(0) = 1$. Under the condition that RWA is valid, $\Omega_{\rm{tg}} \ll \Omega \ll \omega_{\rm{tg}}$, the sensor releases the harmonic response. However, when the condition of RWA is not satisfied, i.e., with large $\Omega_{\rm{tg}}$, small $\omega_{\rm{tg}}$ or small detuning $\xi$ in $\omega_{\rm{tg}} = \omega_1 -\omega_{\acute{0}} + \xi$, the sensor offers complex responses that can be analized with a feed-forward NN. 

We demonstrated our method in the following manner: although in reality the training/validation/test data are taken from experimental measurement, their acquisitions with/under the equivalent conditions can be numerically simulated. At every $t_i$ for each measurement, the result is either $0$ or $1$. With the shot times $n \in 1,2, ..., N_m$, the results become $P_i = \sum_{n=1}^{N_m} z_{n;i}/N_m$, where $z_{n;i}$ is from a Bernoulli distribution deriving from the theoretical probability $P_D(t_i)$. The NN has the input from the measured data $\mathbf{X} = \{P_1, P_2, ..., P_{N_p}\}$ and sends the outputs $\mathbf{Y} = \{y_1, y_2\}$ after processing, aiming at the targets $\mathbf{A} = \{a_1, a_2\} = \{\Omega_{\rm{tg}}, \xi \}$. The elements $P_i$ in the input data set $\mathbf{X}$ are collected at every time instant $t=t_i$ with shot times $N_m$, in the time interval $[0.5t_0, t_0]$, where we choose the period for the ideal harmonic response as $t_0$.  

In order to obtain the high accuracy for the approximation $F(\mathbf{X}) = \mathbf{Y} \approx \mathbf{A}$, we tune the parameters of the NN and adopt the set to give the high estimation precision in Table \ref{tab1}. Levenberg-Marquardt backpropagation, one of the gradient descent algorithms, is applied to train the NN.
\begin{table}[htbp]
\begin{center}
\caption{}{} 
\begin{tabular}{|c|c|}
\hline
Number of neurons in the input layer $N_p$ & 101 \\
\hline
Number of hidden layers & 5 \\
\hline
Number of neurons in each hidden layer & 40 20 12  6 3\\
\hline
Activation function in each hidden layer & $y = \tanh(x)$ \\
\hline
Activation function in the output layer & $y = x$ \\
\hline
Learning rate & $5\times 10^{-3}$ \\
\hline
Shot times for each expectation value & $100$ \\
\hline
Repetition times for each simulated acquisition & $20$ \\
\hline 
\end{tabular}\par
\bigskip
Parameters for training the NN to predict $\mathbf{Y} = \{y_1, y_2\}$ approximating $\mathbf{A} = \{\Omega_{\rm{tg}}, \xi\}$. With the application of the above set,  the average value of the accuracy is above $0.97$.
\label{tab1}
\end{center}
\end{table}
To build a training/validation/test dataset, $241$ examples in $\Omega_{\rm{tg}}/(2\pi) \in [1, 25]$ kHz multiplied by $11$ examples chosen in $\xi / (2\pi)\in [-0.3, 0.3]$ kHz are extracted. Here, we set the above ranges for the parameters as one example to show the estimation abilities of the NN beyond the harmonic response, i.e., $\Omega_{\rm{tg}} = 2\pi\times1$ kHz and $\xi = 0$ kHz. Of course, one can also change the above ranges for the estimation parameters. In addition, the precision of a NN is not sensitive to the choice of the activation function. In the work conducted in \cite{NN-magnetometry}, we consider $N_m=100$ shot times for each expectation value. The finite measurement times gives rise to shot noise, bringing statistical fluctuations. To this end, $20$ times is repeated for each simulated acquisition. Therefore, the total training/validation/test dataset contains $241\times11\times 20$ examples, $70\%$, $15\%$, $15\%$ from which consist of the training, validation and test sets. 

With the presence of shot noise, the estimation of $\Omega_{\rm{tg}}$ and $\xi$ behave in different manners. While the accuracy of the estimation of $\Omega_{\rm{tg}}$ keeps high, it is more difficult for the NN to learn $\xi$ at small $\Omega_{\rm{tg}}$. We divide the whole dataset into the following intervals: $\Omega_{\rm{tg}} / (2\pi) \in [1, 25]$ kHz, $[3.4, 25]$ kHz, $[8.2,25]$ kHz, $[13,25]$ kHz, $[17.8,25]$ kHz and $[22.6, 25]$ kHz, but keeping the same range for $\xi / (2\pi) \in[-0.3, 0.3]$ kHz. For all the above ranges, the correlation coefficient $R$ measures the linear dependence between the outputs of the NN and the targets. The linear regression of the outputs $y$ as the function of the targets $y= \alpha a + \beta$ (with the targets $a$) is compatible with the perfectly matching linear relation $y = a$. When the NN provides better estimation, the outputs approach the targets, $R \rightarrow 1$, and $\alpha \rightarrow 1$, $\beta \rightarrow 0$. In Table \ref{tab2}, NN training results illustrated by $R$, $\alpha$ and $\beta$ are listed in different intervals of $\Omega_{\rm{tg}} / (2\pi)$ with $\xi / (2\pi) \in [-0.3, 0.3]$ kHz.
\begin{table}[htbp]
\begin{center}
\caption{}
\begin{tabular}{|c|c|c|c|}
\hline
$\Omega_{\rm{tg}} / (2\pi) $ $($kHz$)$ & $R$ & $1-\alpha$ & $\beta$\\
\hline
$[1, 25]$ & $0.99356$ & $10^{-2}$ & $6 \cdot 10^{-3}$ \\
\hline
$[3.4, 25]$ & $0.99863$ & $4\cdot 10^{-3}$ & $3\cdot 10^{-3}$   \\
\hline
$[8.2, 25]$ & $0.99972$ & $7\cdot 10^{-4}$  & $3\cdot 10^{-4}$  \\
\hline
$[13, 25]$ & $0.99984$ & $3\cdot 10^{-4}$ & $2\cdot 10^{-4}$  \\
\hline
$[17.8, 25]$ & $0.99996$ &$10^{-4}$  & $4\cdot 10^{-5}$ \\
\hline
$[22.6, 25]$ & $0.99999$ & $7\cdot 10^{-5}$  & $2\cdot 10^{-5}$  \\
\hline
\end{tabular}\par
\bigskip
NN training results illustrated by $R$, $\alpha$ and $\beta$ in different intervals of $\Omega_{\rm{tg}} / (2\pi)$ with $\xi / (2\pi) \in [-0.3, 0.3]$ kHz. The correlation coefficient $R$ measures the linear dependence between the outputs from the NN and the targets, while the linear regression of the outputs is expressed in the function of the targets $y = \alpha a + \beta$. 
\label{tab2}
\end{center}
\end{table}
From Table \ref{tab2}, we can see higher accuracy that the outputs approach the targets is obtained when $\Omega_{\rm{tg}}$ is larger. 
Near the harmonic response regime, i.e. $\Omega_{\rm{tg}} = 2\pi \times 1$ kHz, it is more difficult for the NN to learn different values of $\xi$, as the fluctuation of the responses caused by shot noise can even exceed the difference of the ideal sensor response obtained from the theoretical model. In fact, to distinguish responses from shot noise close to the harmonic regime is difficult, independently of the technique adopted.

One may choose the dataset with the range $\Omega_{\rm{tg}}/(2\pi) \in [8.2,25]$ kHz and $\xi /(2\pi) \in [-0.3, 0.3]$ kHz, randomly pick up $N$ examples with target parameters outside the dataset and check the outputs from the NN. The average value of the accuracy $F_i = \frac{1}{N} \sum_{j=1}^N  |y_i^j - a_i^j|/a_i^j, (i=1, 2) $ is above $0.97$. Of course, one can detect responses at longer time or increase the shot times $N_m$ to improve the detection. However, more energy and time cost with more experiment resources should be paid. To make a balance between the gains and the cost is always needed.  
We demonstrate In Ref. \cite{NN-magnetometry} that in a $^{171}$Yb$^+$ atomic sensor, adequately trained neural networks enable to estimate the parameters of the external field in regimes presenting complex responses under the shot noise due to a finite number of measurements.

\section{Improved quantum perceptron with STA} \label{section:Q-perceptron}
In Ref. \cite{quantum-perceptron}, a quantum perceptron labelled by the subscript $j$ offering the nonlinear response to an input field is constructed via a unitary transformation, written as
\begin{eqnarray}
 \label{gate}
 \hat{U}_j(\hat{x}_j; f) |0_j\rangle = \sqrt{1-f(\hat{x}_j)} |0_j\rangle + \sqrt{f(\hat{x}_j)} |1_j\rangle.
\end{eqnarray}
This transformation is motivated by the {\itshape resting} and {\itshape active} states of a classical neuron that in the quantum description are encoded by the ground $|0\rangle$ and excited $|1\rangle$ states of a qubit. The sigmoidal activation potential of the neuron is codified by the excitation probability of the qubit $f(\hat x_j)=|\langle 1_j |\hat U(\hat x_j;f)|0_j\rangle|^2$. In a feed-forward network, the perceptron gate is conditioned on the field generated by neurons in earlier layers, $\hat{x}_j = \sum_{i=1}^k w_{ji}\hat\sigma^z_i - b_j$, with similar weights $w_{ji}$ and biases $b_j$ as classical networks.

We construct this perceptron gate evolving a qubit with the Ising Hamiltonian 
\begin{eqnarray}
\label{H}
\hat{H} &=& \frac{1}{2}\left[\hat{x}_j  \hat{\sigma}^z_j + \Omega(t)\hat\sigma_j^x\right]
\\
& = & \frac{1}{2} \left[ \sum_{i=1}^k (w_{ji} \hat{\sigma}^z_i \hat{\sigma}^z_j) -b_j \hat{\sigma}^z_j + \Omega(t) \hat{\sigma}_j^x   \right] \nonumber.
\end{eqnarray} 
The qubit is controlled by an external transverse field $\Omega(t)$, has a tuneable energy gap and interacts with other neurons through $\hat{x}_j.$
The Hamiltonian Equation (\ref{H}) has the reduced ground state
\begin{eqnarray}
\label{eigenstate}
|\Phi(x_j/ \Omega(t)) \rangle= \sqrt{1- f(x_j / \Omega(t))} |0\rangle + \sqrt{f(x_j / \Omega(t))} |1\rangle,
\end{eqnarray}
with a sigmoid-like excitation probability of the form
\begin{eqnarray}
\label{sigmoid}
f(x) = \frac{1}{2} \left(1+\frac{x}{\sqrt{1+x^2}}\right).
\end{eqnarray}
Originally, the final state\ \eqref{eigenstate} is achieved in three steps: (i) set the perceptron to the superposition $|+\rangle=\mathcal{H}|0\rangle=\frac{1}{\sqrt{2}}(|0\rangle+|1\rangle)$ with a Hadamard gate; (ii) instantaneously boost the magnetic field $\Omega(0)=\Omega_0\gg |\hat{x}_j|$; (iii) adiabatically ramp-down the transverse field $\Omega(t_f)=\Omega_f$ in a time $t_f$, to do the transformation $\mathcal{A}(\hat{x}_j)|+\rangle\simeq |\Phi(\hat{x}_j/\Omega_f)\rangle$. However, a faster dynamics is always desirable, as a shorter operation time to construct the perceptron response (\ref{sigmoid})  can enhance the performance and reduce the decoherence in the quantum registers. Inverse engineering (IE) methods \cite{STA1} can accelerate the dynamics providing different sigmoid-like excitation probabilities of the qubit leading to an smoother external driving profile $\Omega(t)$ that favors its experimental realization. Moreover, as universality does not depends on the specific shape of the sigmoid-like activation function \cite{Cybenko89}, the resulting improved QNN is still universal \cite{quantum-perceptron-STA}.

More in detail, we parameterize the dynamical state with the undetermined polar $\theta$ and  athimuzal $\beta$ auxiliary angles on the Bloch sphere, 
\begin{eqnarray}
\label{wavefunction}
|\Psi(t)\rangle = \cos(\theta/2) e^{i\beta/2} |0\rangle + \sin(\theta/2) e^{-i\beta/2} |1\rangle.
\end{eqnarray}
Substituting this state $|\Psi(t)\rangle $  into the time-dependent Schr\"{o}dinger equation driven by the Hamiltonian in Equation~(\ref{H}), we obtain that the auxiliary angles satisfy the following coupled differential equations,
\begin{eqnarray}
\label{Omega}
\Omega(t) &=& \dot{\theta}/\sin \beta,
\\
\label{beta}
x_j  &=&  \dot{\theta} \cot\theta \cot\beta - \dot{\beta}.
\end{eqnarray}
Setting the wavefunction  $|\Psi(0)\rangle=|+\rangle$ and $|\Psi(t_f)\rangle=|\Phi(x_j/\Omega_f) \rangle$ at the initial and final times imposes boundary conditions on $\theta(0)$ and $\theta(t_f)$. Moreover, Equation (\ref{Omega}) also imposes similar boundary conditions on $\dot\theta(0)$ at and $\dot\theta(t_f)$ once the initial and final values of the external transverse field $\Omega(0)$ and $\Omega(t_f)$ are specified. 
The set of Eqs. (\ref{Omega}) and (\ref{beta}) are solved by first, interpolating with a polynomial ansatz the function $\theta(t)$ so the boundary conditions of $\theta$ and $\dot\theta$ are fulfilled at $t=0$ and $t_f$, subsequently, deriving $\beta(t)$ from Equation (\ref{beta}). Finally, once $\theta(t)$ and $\beta(t)$ are fully determined, $\Omega(t)$ is deduced from Equation (\ref{Omega}), see \cite{quantum-perceptron-STA} for more details. 
As a result, the initial state $|\Psi(0)\rangle$ is not necessarily the eigenstate of the Hamiltonian avoiding  large $\Omega(0)$ values leading to a smooth and less experimentally demanding $\Omega(t)$ profile compared to fast-quasi-adiabatic implementations \cite{quantum-perceptron}. Moreover, the IE protocol allows a shorter operation time to construct a quantum perceptron and a stable performance with respect to timing errors and the variations of the input potential. In Ref. \cite{quantum-perceptron-STA}, by using STA we propose a speed-up quantum perceptron which has faster performance compared to fast-quasi-adiabatic protocols and enhanced robustness against imperfections in the controls. A deep QNN consisting of a number of our perceptrons can be implemented in physical platforms such as NV center in diamond, trapped ions, and superconducting circuits. One can find the training of such a QNN using gradient descent in the same way as Reference \cite{quantum-perceptron} where the example of searching prime numbers has been given using Python language with quantum perceptron gates evolving adiabatically. The neural potential of each perceptron can include multi-qubit interactions deviating from the current network paradigm of additive activations so that one can  avoid internal hidden layers in a QNN without sacrificing approximative power for information processing tasks, see Ref \cite{multi-NN}.

\section{Conclusion}
In this paper, we introduce the application of NNs on quantum sensing and the development of QNNs by using quantum resources. Particularly, we review the neural-network-based atomic quantum sensor of a $^{171}$Yb$^+$ ion, working in the regime of complex responses beyond harmonic ones. In general, the protocol is applicable to arbitrary quantum detection scenarios. We also review the construction of a quantum perceptron with IE strategies, an efficient method of quantum control, providing faster perceptron performance as well as enhanced robustness against imperfections in the controls. 
Decoherence during the physical implementation in quantum registers can be reduced due to the short operation time of the quantum perceptrons embedded in a QNN.
We hope that more complex quantum problems with different environmental noise will be addressed by NNs, and the performance and training of NNs will be improved by quantum resources.

\section*{Acknowledgment}
We acknowledge financial support from Spanish Government via PGC2018-095113-B-I00 (MCIU/AEI/FEDER, UE), Basque Government via IT986-16, as well as from QMiCS (820505) and OpenSuperQ (820363) of the EU Flagship on Quantum Technologies, and the EU FET Open Grant Quromorphic (828826). J. C. acknowledges the Ramón y Cajal program (RYC2018-025197-I) and the EUR2020-112117 project of the Spanish MICINN, as well as support from the UPV/EHU through the grant EHUrOPE. E. T. acknowledges financial support from Project PGC2018-094792-B-I00 (MCIU/AEI/FEDER,UE), CSIC Research Platform PTI-001, CAM/FEDER Project No. S2018/TCS-4342 (QUITEMAD-CM), and  by Comunidad de Madrid-EPUC3M14.

\end{document}